**D. V. DOBRYCHEVA** [1], Senior Researcher, PhD (Phys. & Math.)
**M. YU. VASYLENKO** [1], Junior Researcher, PhD student
**I. V. KULYK** [1], Senior Researcher, PhD (Phys. & Math.)
**YA. V. PAVLENKO** [1], Head of the Department, Doctor of Sciences
**O. S. SHUBINA** [2], Researcher at the Laboratory, PhD (Phys. & Math.)
**I. V. LUK'YANYK** [3], Head of Department, PhD (Phys. & Math.)
**P. P. KORSUN** [1], Head of the Laboratory, Doctor of Sciences

[1] Main Astronomical Observatory of the National Academy of Sciences of Ukraine
27, Akademika Zabolotnoho Str., Kyiv, 03143 Ukraine
[2] Astronomical Institute of Slovak Academy of Sciences
Tatranská Lomnica, 059 60 Vysoké Tatry, Slovak Republic
[3] Astronomical Observatory of Taras Shevchenko National University of Kyiv
3, Observatorna Str., Kyiv, 04053 Ukraine

*This study introduces an approach to detecting exocomet transits in the dataset of the Transiting Exoplanet Survey Satellite (TESS), specifically within its Sector 1. Given the limited number of exocomet transits detected in the observed light curves, creating a sufficient training sample for the machine learning method was challenging. We developed a unique training sample by encapsulating simulated asymmetric transit profiles into observed light curves, thereby creating realistic data for the model training. To analyze these light curves, we employed the TSFresh software, which was a tool for extracting key features that were then used to refine our Random Forest model training.*

*Considering that cometary transits typically exhibit a small depth, less than 1 % of the star's brightness, we chose to limit our sample to the CDPP parameter. Our study focused on two target samples: light curves with a CDPP of less than 40 ppm and light curves with a CDPP of up to 150 ppm. Each sample was accompanied by a corresponding training set. This methodology achieved an accuracy of approximately 96 %, with both precision and recall rates exceeding 95 % and a balanced F1-score of around 96 %. This level of accuracy was effective in distinguishing between 'exocomet candidate' and 'non-candidate' classifications for light curves with a CDPP of less than 40 ppm, and our model identified 12 potential exocomet candidates. However, when applying machine learning to less accurate light curves (CDPP up to 150 ppm), we noticed a significant increase in curves that could not be confidently classified, but even in this case, our model identified 20 potential exocomet candidates.*

*These promising results within Sector 1 motivate us to extend our analysis across all TESS sectors to detect and study comet-like activity in the extrasolar planetary systems.*

## 1. INTRODUCTION

More than 3700 exoplanets among the 5523 known today have been discovered using the transit method in photometric time series observations of stars conducted by the Kepler, K2, and TESS space missions [3, 12, 31]. The transit method involves detecting periodic decreases in a star's brightness when a planet crosses its disk, blocking part of the radiation in the observer's direction. Typically, exoplanet transits are characterized by periodicity in time and the symmetric dips in stars' light curves.

In addition to the "classical" symmetric patterns of star dimming and subsequent return to normal brightness, typically caused by exoplanets, scientists have identified asymmetric patterns of brightness drops in some stars' light curves. These phenomena have been attributed to the transits of comet-like objects across the stars' disks [13, 17-18, 27, 29, 38, 40]. Such objects possess a central nucleus and an asymmetric, elongated dust atmosphere. The asymmetric transits observed in the photometric light curves are consistent with unusual features previously noted in the absorption spectral lines of young A-class stars. This alignment provides solid evidence for the presence of small bodies in star systems that have debris disks, as discussed in [30].

The studies by [2] and [16] theoretically demonstrate that the evaporation of exocomets, which contain volatile elements and are interspersed with metallic ions like Ca II and Fe II, leads to the creation of planetary atmospheres filled with submicron dust particles. An analysis of approximately 1000 spectra from the star Beta Pictoris revealed around 6000 unique features linked to the transit of comet-like bodies across the star's disk, as reported in [14]. Given that the orbital telescope TESS has already gathered high-precision photometric time series with a 2-minute cadence for over 200 000 stars and continues to do so, the task of identifying and studying comet-like activity within TESS's extensive database is challenging.

The vast amount of data and its rapid accumulation necessitate the use of automatic algorithms for detecting transits and classifying their morphology, primarily through advanced machine learning techniques. Artificial intelligence methods were initially employed on data obtained with the Kepler orbital telescope to classify exoplanet signals and identify false positives, as described in [23]. Studies such as [8, 23, 24] introduced algorithms for verifying exoplanet candidates in the Kepler database and classifying the morphology of star light curves utilizing the classic Random Forest method established in [4]. This strategy proved highly effective, leading to the widespread adoption of deep learning-based algorithms for analyzing data from orbital telescopes, as seen in [1, 26, 33, 34]. A comprehensive review of machine learning applications, encompassing both classical algorithms and deep learning, for the verification of exoplanetary transits, planet classification, and the morphological classification of star light curves in orbital telescope databases is available in works like [24, 32].

In an alternative approach, the authors of [13] developed an automated algorithm specifically designed to search for asymmetrical transits in the Kepler telescope's database. This innovation facilitated the confirmation of previously identified comet transits in the star systems KIC3542116 and KIC11084727 and also led to the discovery of a new transit in the light curves of KIC8027456.

In our study, we introduce an algorithm for detecting exocometary transits in the database of the orbital telescope TESS, utilizing machine learning techniques. Such an approach enables the automated identification of asymmetric transits, thereby substantially decreasing the need for manual visual inspection and expediting the data analysis process. While some of the most sophisticated machine learning methods for planet detection, such as the deep learning approach proposed by Shallue et al. [33], have been successful, we use the classical machine learning method, specifically the Random Forest model. A notable distinction between classical methods and deep learning is that deep learning could autonomously identify features and parameters for sample description. In contrast, classical machine learning requires pre-calculated features as input for the model. It's important to acknowledge that deep learning demands considerable computational resources and does not always allow for the verification of feature significance in a given sample. Upon assessing our available computing resources, we opted to employ classical machine learning methods. We have already presented the preliminary results of our project on detecting asymmetric transits in star light curves collected by TESS and stored in the MAST database using Python code developed by us [36].

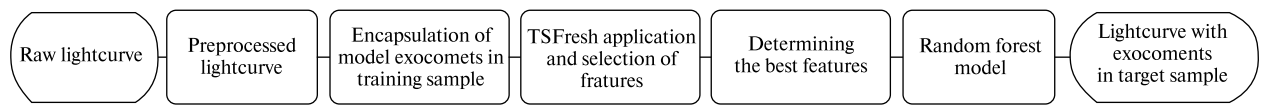

**Figure 1.** Workflow of our method, beginning with raw light curves and culminating in the performance of inference on the sample

This paper is structured into seven sections. Section 2 describes the TESS data utilized in our study. Section 3 details the encapsulation process of simulated asymmetric transit profiles into the observed light curves, thereby preparing the training sample. In Section 4, we introduce the methodology and library employed to extract features from the light curves for subsequent machine learning analysis. Section 5 discusses the machine learning methods applied and the accuracy metrics used for their evaluation. Section 6 presents the outcomes of our research. Lastly, Section 7 offers a concise conclusion of our results. The workflow is shown in Figure 1. More details are given in the following sections.

## 2. TESS DATA

We use a database that was created and is continuously updated based on observations from the TESS space observatory. All light curves observed are stored in the MAST (Mikulski Archive for Space Telescopes) and STScI (Space Telescope Science Institute) data archives. As input data, we use the light curves preprocessed by the operations center team, which have already been corrected for dark current, detector nonlinearity, flat field, scattered background light, focus instability, and other instrumental errors that could lead to artifacts in the light curves [31].

In the archive, the light curves are grouped into sectors, with one sector covering a period of 27 days, during which continuous monitoring of a certain area of the sky is performed. For our study, we use the pre-processed 2-min PDC_SAP normalized light curves. It is worth noting that the short cadence allows for the capture of high-frequency oscillations associated with the physical nature of the stars themselves (pulsations, passage of spots across the star's disk), leading to morphological diversity in the light curves. On the other hand, the relatively large pixel scale (21″ per pixel) introduces a certain dependency of the measured aperture flux on neighboring stars, especially if they are brighter or more variable than the target object.

Therefore, considering these factors, especially the fact that cometary transits have a small depth, less than 1 % of the star's brightness, for this study, we decided to limit the sample by selecting light curves that do not exceed a certain predetermined precision. As a measure of precision, we use the metric of Combined Differential Photometric precision (CDPP), which was developed by the data processing operations center team of the Kepler space telescope and later applied to TESS data processing [7]. CDPP is the signal-to-noise ratio after the smoothing and 'whitening' of light curves, measured in ppm (parts per million) [10, 35]. For example, if a light curve with a CDPP level of 20 ppm has a transit lasting 3 hours with a depth of 20 ppm (or 0.002 %), it is expected that this transit can be detected, as it corresponds to a signal at the level of 1 sigma. This metric is very convenient for assessing the precision of light curves, and it is easy to calculate at every step of the data processing.

## 3. PREPARATION OF THE TRAINING SAMPLE

The machine learning approach suggests building a classifier, which is able to separate data into multiple classes. In order to predict possible transits with the asymmetric shapes in the light curves, we use a binary classifier enabling the separation of light curves into two classes: 'exocomet candidate' and 'non-candidate'. To train the machine learning model, we need to have the training sample containing light curves of both types: with asymmetric transits and "typical" light curves from the database. As there have been few exocomet transits detected so far in the observed light curves, to create the training sample with the "exo-comet candidates", we use the simulated asymmetric transit profiles encapsulating them into the observed light curves. The approach to simulation of the asymmetric transits is described in detail in the following papers [5, 27]. Figure 2 shows the simulated transits normalized to 1 for three different cases of morphologically different cometary comae resembling the solar system comet C/2006 S3, which could cause asymmetric dips in the star brightness when crossing the star's disk. The influence of an impact parameter on the transit shape (the transit through the center and through the edge of the star disk) is depicted in Figure 3. It is seen that the morphology of the transit shapesis different, but all of them are asymmetrical. We simulate about 300 such profiles to encapsulate them into the observed light curves from the first sector of the TESS data. We randomly select both the simulated transit profile and the light curve into which the transit is then inserted. Finally, the training sample contains 20 000 light curves: 10 000 of them are labeled as '0' marking 'non-candidate', and 10 000 have the label '1', or 'exocomet candidate'. In order to study the influence of the light curve precision on the transit prediction score, we prepared two target samples, selecting the light curves with the CDPP parameter less than 40 ppm and 150 ppm, respectively. Finally, we have two kinds of target samples: about 2000 light curves whose CDPP is less than 40 ppm, and about 9000 light curves with the CDPP less than 150 ppm. Each sample is provided with a corresponding training sample prepared as it is described above.

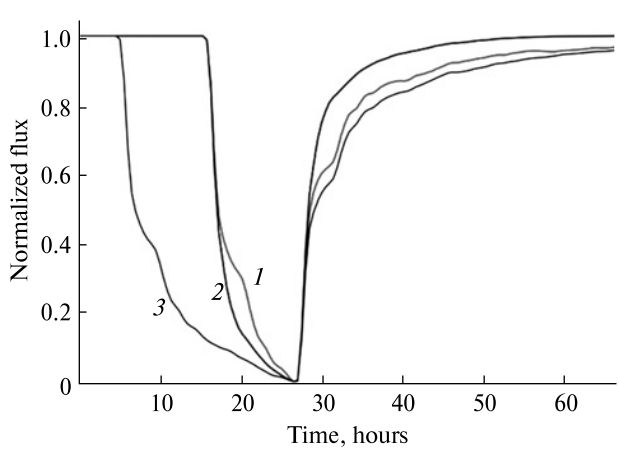

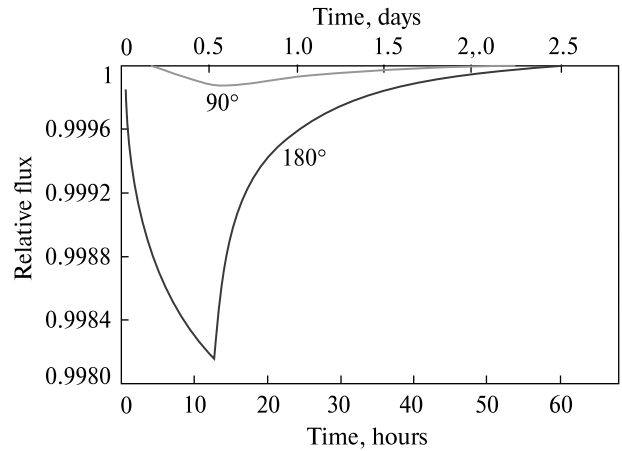

**Figure 2.** Simulated transit profiles for three different cases of morphologically different cometary comae, which cause the exocomet transits

**Figure 3.** Simulated transit profile for different impact parameters: a comet crosses the star disk through its center (180°) and at the star edge (90°)

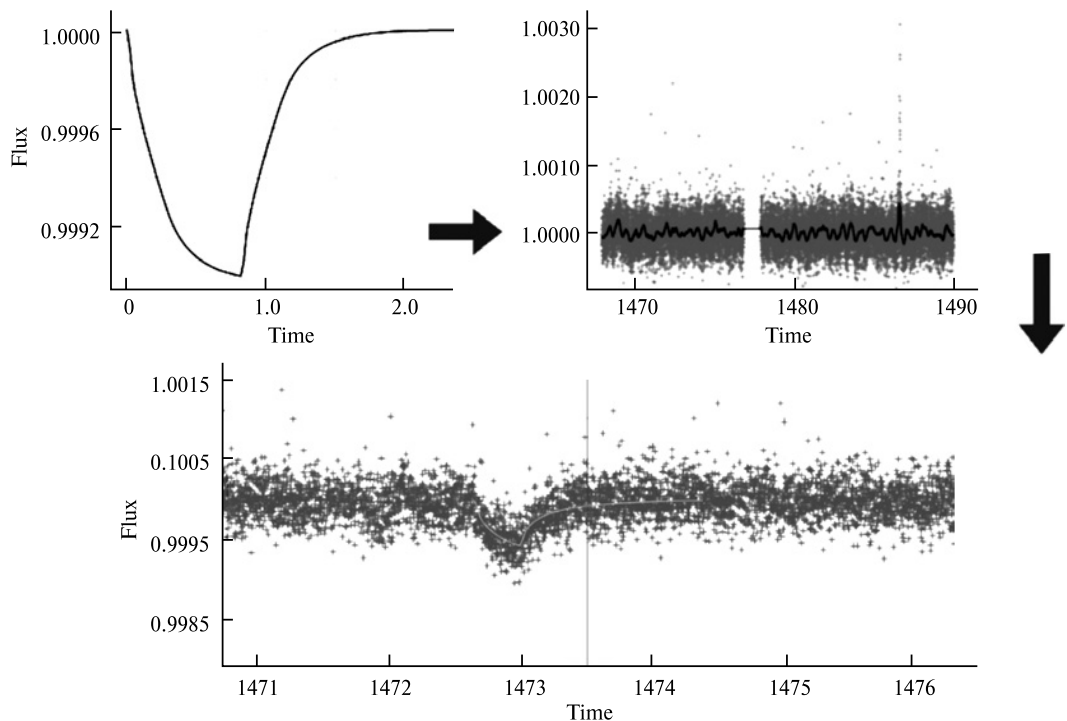

**Figure 4.** Top panel: transit profile (left) and the light curve (right), both selected randomly; bottom panel: the segment of the light curve with the transit

Figure 4 demonstrates the procedure of the transit injection in detail. The top panel shows the transit profile and the light curve both randomlyselected from the corresponding datasets, the bottom panel depicts thesegment of the light curve with the transit injected.

### 4. FEATURE EXTRACTION

After the training sample has been prepared, the next step is feature extraction, which accurately represents statistical information about the characteristics of each light curve. These features are used as the input data for the Random Forest model both for training and predictions. Typically, standard features include standard statistical information, such as minimum and maximum values in a certain interval, the number of the minimums in the light curve, the mean and standard deviation, the number of values that are higher and lower than the mean value, and other statistics. Features can be conventional, as already mentioned, or more unconventional, such as the p-value of the slope coefficient of the trend line in the current moving window. We use the TSFresh library for Python for the feature extraction process. The library provides more than 60 functions to calculate different features, however, the process can be time-consuming [6]. Additionally, not all features are useful, and an excessive number of the features can lead to overfitting. To optimize the parameters, the library provides tools for evaluating the significance of features for regression or classification tasks.

We used the efficient feature extraction settings of TSFresh, resulting in about 790 generalized time series features. We removed all irrelevant features to balance the informativeness of features and computational resource expenditure well. We select a subset of features from the full set that are typically most informative for time series analysis. Finally, about 390 most significant features are used for training the classifier based on the Random Forest method.

### 5. RANDOM FOREST

The Random Forest method is very robust in processing the spectro-phohtometry data as the tool for detachment of various features of exoplanet light curves [22], exoplanetary atmospheres [9, 25], exoplanet prediction [39] and resonant Koiper Belt objects search [19, 20]. We deeply exploited it for galaxy morphological classification [15, 37]. Our binary classifier model is designed to sort light curves into two primary classes: 'exocomet candidate' and 'non-candidate'. For this purpose, we used the Random Forest (RF) model [28], an ensemble of Decision Trees trained on input data. Each tree in this ensemble functions as an independent classifier, and their collective outcomes are integrated to form a final prediction. Throughout its training phase, the Random Forest model undergoes multiple iterations, learning from subsets of data and employing a random selection of features in each iteration. This approach makes the model resistant to overfitting and enhances its ability to generalize effectively to new data.

During the training process, we employed a 5-fold cross-validation technique. This involves dividing the training dataset into five equal segments, or 'folds'. In each of the 5 training cycles, the model is trained using 4 folds, while the fifth fold is reserved for validation. Each fold is used for validation exactly once. The final performance metric of the model is calculated as the average of the performance obtained at each validation stage. This method ensures that every data point is used for both training and validation, enhancing the robustness and reliability of our model.

To evaluate the outcomes of our machine learning model, we employ the following metrics:

• *Accuracy*: this metric represents the proportion of correctly classified samples, encompassing both 'exocomet candidates' and 'non-candidates'. It measures the overall correctness of the model in classification tasks.

• *Precision*: this is the ratio of predicted instances identified as 'exocomet candidates' that are actually true exocomets. Precision focuses on the accuracy of the model's positive predictions but doesn't account for cases where true positives were missed.

• *Recall*: this metric reflects the ratio of actual exocomets correctly identified by the classifier as 'exocomet candidates'. Recall indicates the proportion of actual positive cases that the model successfully detects.

• *F1-score*: this is the harmonic mean of precision and recall. It is a useful metric when you want to balance precision and recall, especially if there's an uneven class distribution or if false positives and false negatives have different costs.

As we previously noted, there are only a limited number of cases where exocomet transits have been successfully detected. Taking TESS mission data as an example, only about 3 % of all light curves contain potential indications of exoplanets [21]. Thus, the data on exocomets is even scarcer. This imbalance in the datasets means that the accuracy metric alone is not sufficiently informative for evaluating the effectiveness of exocomet detection algorithms. In such disproportionate conditions, a classifier predicting the 'non-candidate' category for all light curves could display an impressive accuracy of around 97 %, which can be misleading (like in work [21] about 'planet candidates'). Such high accuracy might not reflect the model's ability to identify potential exocomet candidates, indicating the necessity of employing additional metrics for a more comprehensive performance assessment.

In assessing model accuracy, relying solely on precision can lead to missing many potential candidates. High precision implies that most identified 'exocomet candidates' are likely true candidates. However, the issue arises because high precision can be achieved by making only a few very confident predictions about the presence of 'exocomet candidates' and ensur-ing their correctness. In searching for exocomets, we prefer a strategy tolerating more false positives rather than missing true planetary signals. Thus, recall becomes a much more critical metric in evaluating the effectiveness of our exocomet detection algorithm. It prioritizes capturing true signals over avoiding false alarms, which is essential in the exploratory phase of exocomet discovery.

Another crucial hyperparameter in classification models based on decision thresholds is the balance between precision and recall. This threshold determines at what predicted probability level a light curve is classified as an 'exocomet candidate'. Setting a higher decision threshold leads to an increase in the model's precision at the expense of recall, as the model becomes more conservative in its predictions. Conversely, lowering the threshold enhances recall but may reduce precision. Keeping this in mind, the hyperparameter can be adjusted to achieve an optimal balance, particularly ensuring a high recall. In standard classification tasks, a threshold of 0.5 is commonly used, where light curves with a predicted probability above 50 % are classified as 'exocomet candidates'.

## 6. RESULTS

To obtain the training outcome for the model, each training sample is divided into two: the training sample and the test sample. The primary difference is that the model "sees" the labels of the training sample but can not see the labels of the test sample, which challenges the model to make predictions for both. Figure 5 presents the training outcomes for models developed using training and test subsamples constructed from the 40 ppm or lower precision light curves. The figure illustrates how the accuracy of these predictions correlates with the depth of learning within the model. This 'depth of learning' is a parameter controlling the maximum depth of decision trees during the model training phase. The graph shows two lines: a dashed line for the training sample, showing a trend where accuracy improves as tree depth increases — this is typical, as more complex trees can capture the nuances of the training sample more effectively. In contrast, the solid line for the test sample highlights the need for precise tree depth calibration to avoid overfitting — where the model be-comes too specialized for the training sample — and underfitting — where the model is too simplistic to grasp the data's complexity. Table 1 presents the corresponding performance metrics.

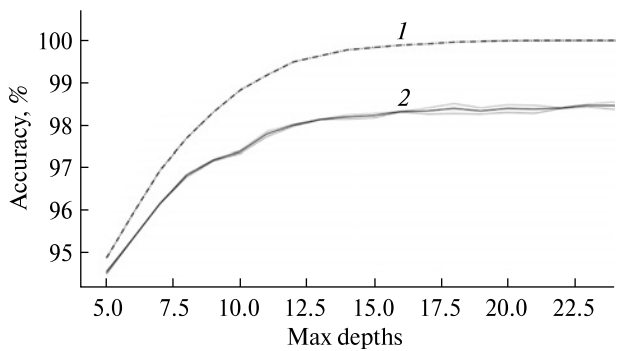

***Figure 5.*** Accuracy of model predictions as a function of learning depth for the training sample constructed for a subset of light curves with CDPP ≤ 40 ppm: curve *1* — training data, *2* — test data

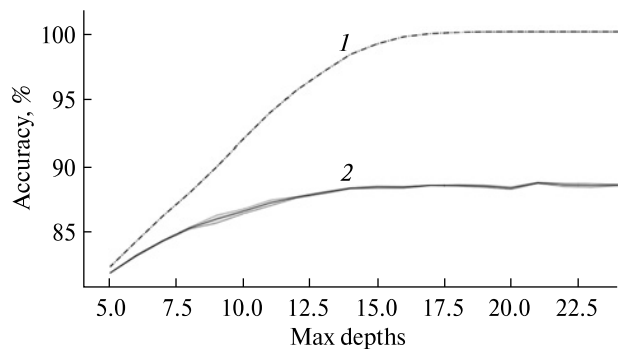

***Figure 6.*** Accuracy of model predictions as a function of learning depth for the training sample constructed for a subset of light curves with CDPP ≤ 150 ppm: curve *1* — training data, *2* — test data

*Table 1.* **Classification reports on the Random Forest classifier applied on the training sample for sector 1 with a CDPP of 40 ppm, featuring 20 000 light curves (10 000 comet transits, 10 000 — randomly selected light curves)**

|  | Precision | Recall | F1-score | Support |
|---|---|---|---|---|
| 0 | 0.9832 | 0.9537 | 0.9682 | 1600 |
| 1 | 0.9550 | 0.9837 | 0.9692 | 1600 |
| Accuracy |  |  | 0.9687 | 3200 |
| Macro avg | 0.9691 | 0.9687 | 0.9687 | 3200 |
| Weighted avg | 0.9691 | 0.9687 | 0.9687 | 3200 |

*Table 2.* **Classification reports on the Random Forest classifier applied on the training sample for sector 1 with a CDPP of 150 ppm, featuring 20 000 light curves (10 000 comet transits, 10 000 — randomly selected light curves)**

|  | Precision | Recall | F1-score | Support |
|---|---|---|---|---|
| 0 | 0.7811 | 0.9690 | 0.8649 | 2000 |
| 1 | 0.9591 | 0.7285 | 0.8280 | 2000 |
| Accuracy |  |  | 0.8487 | 4000 |
| Macro avg | 0.8701 | 0.8487 | 0.8465 | 4000 |
| Weighted avg | 0.8701 | 0.8487 | 0.8465 | 4000 |

Table 1 presents the classification report on the Random Forest classifier applied to the training sample with a CDPP threshold of 40 ppm for sector 1, including 20 000 light curves (10 000 with comet transits and 10,000 randomly selected). The classifier exhibits impressive precision and recall metrics across both classes (0 for 'non-candidate' and 1 for 'exocomet candidate'), with scores above 0.95. This high accuracy (0.9687) is noteworthy, indicating the model's robust ability to identify exocomets transit correctly within light curves. The balanced F1-scores, hovering around 0.968, further confirm the model's proficient classification capabilities, suggesting a finely-tuned algorithm that maintains a calibrated equilibrium between precision and recall.

Such high-performance metrics underscore the potential of the Random Forest approach in reliably detecting exocometary transits in the sample provided.

As noted above, we prepared two samples depending on the light curves' accuracy. Figure 6 shows the accuracy of predictions as a function of the model's depth of learning, and it evidences a decline in classifier performance upon escalating the CDPP threshold to 150 ppm. The decrease in the predictive probability is also seen in Table 2, where the classification report for the same classifier but with a higher CDPP threshold of 150 ppm is listed. Notably, there is a significant reduction in precision for 'non-candidate' curves (class 0), with a value of 0.7811, and a slight decrease in recall for 'exocomet candidate' curves (class 1) to 0.7285. These findings indicate a deterioration in the classifier's capacity to effectively discern transits, corroborated by an overall accuracy reduction to 0.8487.

The macro and weighted averages of precision and recall present a similar picture, indicating a more challenged classifier at this threshold. This reduction in performance metrics, particularly in the recall of class 1, highlights the increased difficulty in detecting transits. The reason for the decrease in the predictive probability is due to 1) a decrease in the signal-to-noise ratio of the input light curves and 2) a lower accuracy threshold leads to expanding the input sample (about 9000 compared to 2000) and, as a result, including light curves with amore diverse morphology.

Figure 7 illustrates the classification outcomes for the training samplewhen subjected to a CDPP threshold of 40 ppm. The figure distinctly showcases the model's capacity for distinguishing between light curves with asymmetric transits (probability estimate is more than 0.6) and without asymmetric ones. Figure 8 provides an assessment of the classifier's performance on the target sample at the same CDPP threshold of 40 ppm. It reveals that, generally, the likelihood of detecting asymmetric transits is low across the sample (the probability estimate is less than 0.5 for most light curves). However, there are notable exceptions where 12 light curves demonstrate high predictive probabilities. This suggests that despite a predominant low detection rate, the model has identified a subset of light curves with a strong likelihood of exhibiting exocomet transits, affirming the potential of the Random Forest approach in flagging significant transit events amidst a vast data array.

Figure 9 shows the training sample with a CDPP threshold of 150 ppm, and we can see highlighting the challenges in differentiating light curves. The graphic shows a more blended distribution of light curves, suggesting that increasing the CDPP threshold to 150 ppm reduces the classifier's effectiveness. But still, it is clearly visible that the method selects those light curves in which there are asymmetric transits.

Figure 10 provides the histogram of the probability estimates for the target sample at the same CDPP threshold of 150 ppm. You can see that the majority of light curves cannot be confidently categorized, having probability estimates between 0.2 and 0.8, though 20 light curves exhibit high probabilities to have asymmetric transits.

Our investigation into the detection of asymmetric exocomet transits ('exocomet candidate') within the TESS database, utilizing the Random Forest method, reveals the impact of the CDPP threshold on classifier performance. At the lower threshold of 40 ppm, the model demonstrates excellent capability in distinguishing light curves with potential asymmetric transits ('exocomet candidate'), as substantiated by the precision and recall metrics that both exceed 0.95 for 'exocomet candidate'. The F1-score, harmonizing these metrics, corroborates the classifier's adeptness, maintaining scores near 0.97 for both classes ('exocomet candidate' and 'non-candidate'), suggesting a well-calibrated balance between precision and recall.

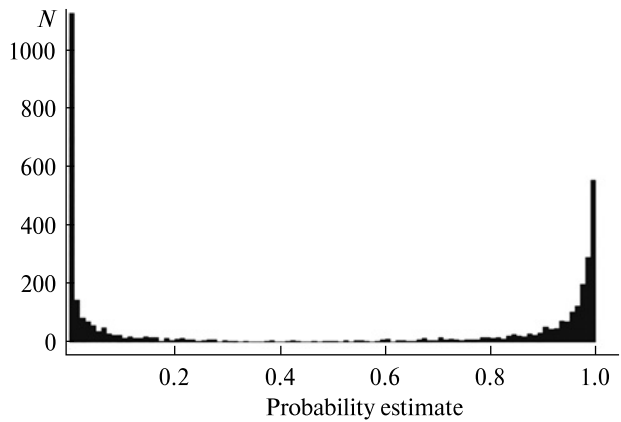

***Figure 7.*** Histogram of the Random Forest classifier's output probabilities for the training sample (CDPP threshold is 40 ppm)

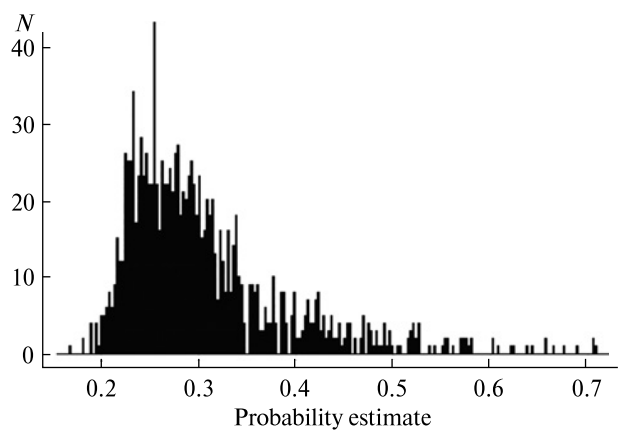

***Figure 8.*** Histogram of the Random Forest classifier's output probabilities for the 2000 light curves of the target sample (CDPP threshold is 40 ppm)

The CDPP threshold to 150 ppm, however, introduces significant challenges. The classifier's pre-cision for 'non-candidate' curves weakens, and the recall for 'exocomet candidate' curves suffers, indicating a lessened ability to detect transits without incurring false negatives. This is particularly evident in the target sample analysis, where most curves reside within an indeterminate range, reflecting the classifier's struggle to make definitive classifications under this more conservative threshold.

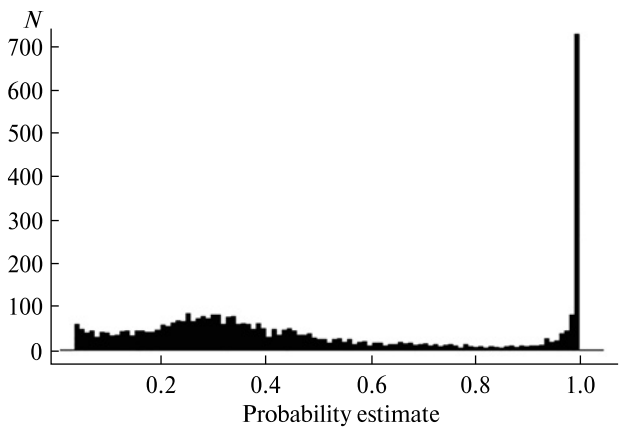

***Figure 9***. Histogram of the Random Forest classifier's output probabilities on the training sample, applying a CDPP threshold of 150 ppm

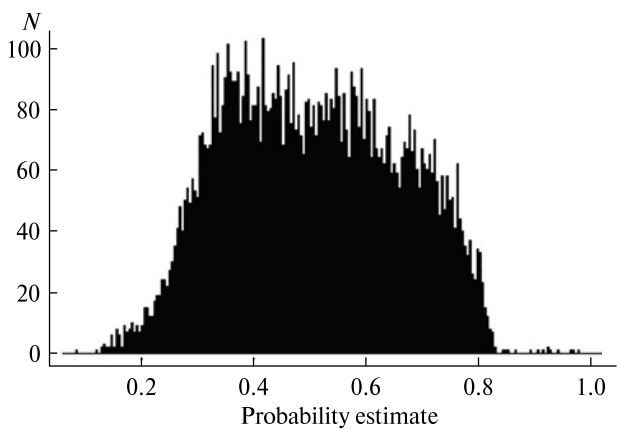

***Figure 10***. Histogram of the Random Forest classifier's output probabilities for the 9000 light curves of the target sample (CDPP threshold is 150 ppm)

The findings underscore the critical role of the CDPP threshold in the search for 'exocomet candidate'. While a lower CDPP threshold enables more precise identification of potential asymmetric transits, enhancing the prospect of discovery, a higher threshold risks overlooking genuine asymmetric transit signals, although potentially reducing false positives. These insights provide a compelling case for the meticulous optimization of CDPP thresholds in the broader context of exoplanet and exocomet transit detection.

Our study demonstrates the potential of the Random Forest method in efficiently screening large datasets for exocomet transits. The careful selection and optimization of CDPP thresholds are crucial to maximizing the effectiveness of these machine learning techniques. These insights provide a valuable framework for future explorations in exoplanet and exocomet transit detection using TESS data.

## 7. CONCLUSIONS

This study shows the effectiveness of machine learning, in particular the Random Forest algorithm, on detecting potential exocomet transits in the TESS Sector 1 data, which significantly reduced the need for manual inspection of light curves. Given the limited number of exocomet transits detected in the observed light curves, creating a sufficient training sample for the machine learning method was challenging. To address this, we generated a realistic training sample by incorporating simulated asymmetric transit profiles into observed light curves. We then utilized TSFresh software to extract key features from these curves.

Considering that cometary transits typically exhibit a small depth, less than 1 % of the star's brightness, we chose to limit our sample to the CDPP parameter, which is a measure of the signal-to-noise ratio after smoothing and 'whitening' the light curves. Our study focused on two target samples: about 2000 light curves with a CDPP of less than 40 ppm and about 9000 light curves with a CDPP of up to 150 ppm. Each sample was accompanied by a corresponding training set.

This methodology achieved an accuracy of approximately 96 %, with both precision and recall rates exceeding 95 % and a balanced F1-score of around 96 %. This level of accuracy was effective in distinguishing between 'exocomet candidate' and 'non-candidate' classifications for light curves with a CDPP of less than 40 ppm. From the sample of about 2000 light curves, our model identified 12 po-tential exocomet candidates (with a CDPP of less than 40 ppm) and 20 from the larger sample of approximately 9000 light curves (with a CDPP of up to 150 ppm). However, when applying machine learning to less accurate light curves (up to 150 ppm), we noticed a significant increase in curves that could not be confidently classified. This was due to the larger sample size (about 9000 compared to 2000) and the increased diversity in light curve morphology.

For the further effective application of machine learning for TESS data, we need to implement an effective technique that does not require large computing resources, cleaning the brightness variations from fluctuations associated with the star itself. These promising results from Sector 1 serve as a significant motivation for us to extend our analysis across all TESS sectors, thereby broadening the scope and potential impact of our research in the realm of astronomical studies.

**Acknowledgments.** *This study was performed in the frame of the government funding program for institutions of the National Academy of Sciences of Ukraine (NASU) and supported by the National Research Foundation of Ukraine (№ 2020.02/0228). All data presented in this paper were obtained from the Mikulski Archive for Space Telescopes (MAST). STScI is operated by the Association of Universities for Research in Astronomy, Inc., under NASA contract NAS5-26555. Support for MAST for non-HST data is provided by the NASA Office of Space Science via grant NNX09AF08G and by other grants and contracts. This paper includes data collected by the TESS mission. Funding for the TESS mission is provided by the NASA Science Mission directorate.*

*Д. В. Добричева*[1], старш. наук. співроб., канд. фіз.-мат. наук

*М. Ю. Василенко*[1], молодший наук. співроб.

*І. В. Кулик*[1], старш. наук. співроб., канд. фіз.-мат. наук, старш. дослід.

*Я. В. Павленко*[1], зав. відділу, д-р фіз.-мат. наук, старш. дослід.

*О. С. Шубіна*[1,2], наук. співроб., канд. фіз.-мат. наук

*І. В. Лук'яник*[3], зав. сектору, канд. фіз.-мат. наук, старш. дослід.

*П. П. Корсун*[1], зав. лаб., д-р фіз.-мат. наук, старш. дослід.

[1] Головна астрономічна обсерваторія Національної академії наук України

вул. Академіка Заболотного 27, Київ, Україна, 03143

[2] Астрономічний Інститут Словацької академії наук

Високі Татри, Татранська Ломниця, Словацька Республіка, 059 60

[3] Астрономічна обсерваторія Київського національного університету імені Тараса Шевченка

вул. Обсерваторна 3, Київ, Україна, 04053

## ПОШУК ТРАНЗИТІВ ЕКЗОКОМЕТ У БАЗІ ДАНИХ TESS ЗА ДОПОМОГОЮ МЕТОДУ ВИПАДКОВОГО ЛІСУ

В даному дослідженні представлено ефективний підхід до виявлення екзокометних транзитів у даних першого сектору космічного телескопу Transiting Exoplanet Survey Satellite (TESS). Враховуючи обмежену кількість наявних екзокометних транзитів, що виявлені у спостережуваних кривих блиску, створення репрезентативної тренувальної вибірки для машинного навчання є великою складністю. Ми розробили унікальну тренувальну вибірку шляхом інкапсуляції змодельованих асиметричних профілів транзиту в спостережувані криві блиску, таким чином створюючи реалістичні дані для навчання моделі. Щоб проаналізувати ці криві блиску, ми використали програмний пакет TS-Fresh, який служив інструментом для виявлення ключових ознак, які потім використовувалися для вдосконалення нашої моделі Випадковий ліс при навчанні.

Враховуючи, що кометні транзити зазвичай мають невелику глибину, менше ніж 1 % яскравості зірки, ми вирішили обмежити вибірку параметром CDPP. Наше дослідження було зосереджено на двох цільових вибірках: криві блиску з CDPP менше 40 ppm та криві блиску з CDPP до 150 ppm. Кожана вибірка супроводжувалася відповідною тренувальною вибіркою. Наш метод продемонстрував високу точність, досягнувши показника біля 96 %, в поєднанні з високими показниками влучності (Precision) та повноти (Recall) для обох, які перевищують 95 %, а також збалансованими показниками F1-міри на рівні 96 %. Цей рівень точності відповідає ефективному розпізнаванню транзитів 'кандидат на екзокомету' та 'не кандидат' для кривих блиску з CDPP менше 40 ppm, при цьому наша модель ідентифікувала 12 потенційних екзокомет-кандидатів. Однак, застосовуючи машинне навчання до менш точних кривих блиску в яких CDPP до 150 ppm, ми помітили значне збільшення кривих, які не можна було впевнено класифікувати, але навіть у цьому випадку наша модель ідентифікувала 20 потенційних екзокомет-кандидатів.

Ці багатообіцяючі результати в першому секторі спонукають нас розширити наш аналіз на всі сектори TESS для виявлення та вивчення кометоподібної активності в позасонячних планетарних системах.

***Ключові слова:*** комети, планетні системи, малі планети; затемнення, транзити, планети та супутники.